\documentclass{aastex63}
\usepackage{graphics}
\usepackage{pgfplotstable}
\usepackage{float}
\makeatletter
\submitjournal{APJ}

\shorttitle{Cosmological evolution on Short GRBs}
\graphicspath{{./}{figures/}}
\pgfplotsset{compat=1.6}
\begin{document}

\title{Cosmological Evolution of the Formation Rate of Short Gamma-ray Bursts With and Without Extended Emission}

\correspondingauthor{Dainotti, M. G.}
\email{maria.dainotti@nao.ac.edu.jp}

\author{Dainotti, M. G.}
\affiliation{National Astronomical Observatory of Japan; Mitaka, Tokyo, Japan; mdainott@stanford.edu}
\affiliation{Space Science Institute, Boulder, Colorado}
\author{Petrosian, V.}
\affiliation{Department of Physics, Stanford University, Via Pueblo Mall 382, Stanford, CA 94305-4060, USA;vahep@stanford.edu}
\affiliation{Kavli Institute of Particle Astrophysics and Cosmology, Stanford University}
\affiliation{Department of Applied Physics, Stanford University}
\author{Bowden, L.}
\affiliation{Cornell University, Ithaca, New York, USA}

\begin{abstract}
Originating from neutron star-neutron star (NS-NS) or neutron star-black hole (NS-BH) mergers, short gamma-ray bursts (SGRBs) are the first electromagnetic emitters associated with gravitational waves. This association makes the determination of SGRB formation rate (FR) a critical issue. We determine the true SGRB FR and its relation to the cosmic star formation rate (SFR). This can help in determining the expected Gravitation Wave (GW) rate involving small mass mergers. We present non-parametric methods for the determination of the evolutions of the luminosity function (LF) and the FR using SGRBs observed by {\it Swift}, without any assumptions. These are powerful tools for small samples, such as our sample of 68 SGRBs. We combine SGRBs with and without extended emission (SEE), assuming that both descend from the same progenitor. To overcome the incompleteness introduced by redshift measurements we use the Kolmogorov-Smirnov (KS) test to find flux thresholds yielding a sample of sources with a redshift drawn from the parent sample including all sources. Using two subsamples of SGRBs with flux limits of $4.57 \times 10^{-7}$ and  $2.15 \times 10^{-7}$ erg cm$^{-2}$ s$^{-1}$ with respective KS {\it p=(1, 0.9)}, we find  a 3 $\sigma$ evidence for luminosity evolution (LE), a broken power-law LF with significant steepening at $L\sim 10^{50}$ erg s$^{-1}$, and a FR evolution that decreases monotonically with redshift (independent of LE and the thresholds). Thus, SGRBs may have been more luminous in the past with a FR delayed relative to the SFR as expected in the merger scenario. 
\end{abstract}

%In this analysis, we consider short GRBs, short GRBs with extended emission and intrinsically short (GRBs that have the intrinsic duration of the prompt emission $< 2$ s in the rest frame).


\section{Introduction} \label{sec:intro}
The bi-modality of the distributions of the duration of the prompt emission of gamma-ray bursts (GRBs) that separates them into two classes, short and long (hereafter SGRBs and  LGRBs), at the observer frame duration $T^{obs}_{90}\sim 2$ s
(\cite{K}),
\footnote{$T_{90}$ is the time in which a burst emits from $5\%$ to $95\%$ of its total measured counts.}
remains valid for rest frame duration $T_{90}=T^{obs}_{90}/Z$ after measurement of redshift $Z=z+1$.%
\footnote{The quantity $Z=1+z$ is more convenient than $z$ for describing the evolutionary functions at high redshifts. We refer to both variables as redshift.}  
In addition, SGRBs tend to have harder spectra, and are located in the outskirts of older host galaxies rather than in star-forming galaxies with a younger stellar population (\cite{F,Wa}). These differences have led to two separate progenitors: the collapse of massive stars (\cite{W,MF}) for LGRBs and the merger of two neutron stars (NSs) or a NS and a black hole (BH)  for SGRBs (\cite{LS,E,N,Na,Ber}).

The possibility and eventual discovery of GW radiation from several BH-BH mergers (\cite{A2,A3,A4}) and one NS-NS merger (GW 170817, \cite{A5}, which is associated with the SGRB 170817A, has 
made the determination of the intrinsic distributions, such as the LF, $\Psi(L,Z)$, luminosity evolution $L(Z)$, and formation rate (co-moving density) evolution $\dot \rho(Z)$ of SGRBs a critical issue (see e.g.~ \cite{WP,T, ZW, Gh3, P}). 

Categorizing GRBs into long and short classes is not straightforward, and some sub-classes exist: for example,  SGRBs followed by a low flux long extended  emission (\cite{NB}, hereafter SEE). The nature of SEEs is still being debated (\cite{D10,Y}), and there are arguments (\cite{BP}) in favor of them having the same progenitors (merger events) as the usual SGRBs. 
Thus, we consider SGRBs and SEEs together in a combined sample. 
We aim to obtain a more robust determination of the above mentioned intrinsic distributions using non-parametric (instead of commonly used forward fitting that involve several assumptions) methods, described in \S 2, and a selected sample of SGRBs with measured spectroscopic redshifts, described in \S 3.
The results are presented in \S 4 followed by a summary and conclusion section.

\section{The Methodology}
\label{methods}

Determination of the intrinsic distributions requires a sample with well defined observational selection criteria, such a defined flux limit referred to as a {\it reliable sample}. The most readily available ``reliable" samples are those with a well defined detection threshold or energy flux limit, $f>f_{\rm lim}$. For GRBs, one also requires redshift to obtain the luminosity and its truncation:
\begin{equation}
\label{Lz}
L_i(Z)=4\pi d_L^2(Z, \Omega)f_iK(Z,\alpha),\,\,\,\, {\rm for} \,\,\,\,  f_i=f\,\,\, {\rm and} \,\,\, f_{\rm lim},
\end{equation}
respectively. Here $K(Z, \alpha)$ are the K-correction, where $\alpha=d\ln f/ d\ln \nu -1$ is the photon number spectral index. We use energy fluxes, and hence luminosities, integrated over the {\it Swift} energy band $15-150$ keV. 
To compute $K$ we used a power-law with exponential cutoff spectrum, that fits best to most GRBs, with the best fit values  taken from the online Third GRB Catalog (Lien et al. 2016). For 9 cases for which the cutoff power law was not a possible fit we used the simple power law. The derived luminosities are shown in right panel of Fig. 2.%
\footnote{Here $d_L$ is the luminosity distance using the Hubble constant $H_0=0.70$ km $^{-1}$ s$^{-1}$ Mpc$^{-1}$, and  density parameters $\Omega_m=0.3$ and $\Omega_\Lambda=0.7$.} 
These information are used to determine the bi-variate distribution $\Psi(L,Z)$ taking account the bias (the Malmquist bias (1922)) introduced by the flux limit.
A common practice to account for this bias is to
use some forward fitting method, whereby a set of assumed {\it parametric functional forms} are fit to the  data to determine the ``best fit values" of the many parameters of the functions, raising questions about the uniqueness of the results.

Non-parametric, non-binning methods,
such as the so-called $V/V_{\rm max}$ method (Schmidt 1968) and the $C^{-}$ method of Lynden-Bell (1971),  
require no such assumptions and are more
powerful, especially for small samples.  However, as pointed out by Petrosian (1992), these methods require the critical assumption that the variables, in this case $L$ and $Z$, are uncorrelated, which implies the physical assumption of no luminosity evolution ({\bf LE}); i.e.~$\Psi(L,Z)=\phi(L){\dot \rho}(Z)$).
This shortcoming led to developments of the more powerful (also non-parametric, non-binning) methods of Efron and Petrosian (1992, 1999), which does away with the no-LE assumption by testing whether $L$  and $Z$ are  correlated. If correlated then it introduces a new
variable $L_0\equiv L/g(Z)$ and finds the LE
function, $g(Z)$, that yields an uncorrelated $L_0$ and $Z$. 
For normalization $g(Z=1)=1$, $L_0$ is the local $z=0$ luminosity. Thus, the LF reads as 
\begin{equation}
\label{LF}
\Psi(L,Z)=\frac{\dot \rho(Z)}{g(Z)}\phi\Big[\frac{L}{g(Z)},\alpha_i\Big],
\end{equation}
where $\alpha_i$ is the shape parameters.%, e.g. power-law index.%
\footnote{For a small sample of sources determination of the evolution of the shape parameters is difficult to obtain. Here we assume that they are independent of the redshift.}
One can then proceed with the determination of the local luminosity function $\psi(L_0)$ and density rate evolution $\dot \rho (Z)$.
This combined Efron-Petrosian and Lynden-Bell (EP-L) method has been
very useful for studies of evolution of GRBs (Petrosian et al. 2015).
Thus, more papers dealing with GRB evolution use this method. Recent analyses of different LGRB samples  (Petrosian et al.~2015; Yu et
al.~2015; Pescalli et al.~2016; Tsvetkova et al.~2017; Lloyd-Ronning et al. 2019) show similar results, indicating that, contrary to the common assumptions, there is a {\bf significant LE}, and that there is a considerable disagreement between LGRB FR and the SFR at low redshifts (${\dot \rho(z)}>SFR$ for $z<1$). A similar high formation rate of SGRBs at low redshifts will have
profound consequences on the expected rate of GW sources.
There have been several analyses of SGRBs (see, e.g.~Wanderman \& Piran 2015: Ghirlanda et al.~2016; Paul 2017; Yonetoku et al. 2014; Zhang \& Wang 2018). The last two papers use the EP-L method and so-called pseudo redshifts for samples of
45 and 239 sources, respectively, with somewhat different results.

We note, however, that many aspects of  determination of  a true  LF, and its evolution continue to be debated. In particular, the unique aspect of the EP-L method, namely the determination of  correlation between luminosity and redshift (i.e. the luminosity evolution, LE) is sensitive to the flux threshold; using a lower threshold can lead to a stronger luminosity evolution.  Thus, in our past work on LGRBs and here we evaluate the evolution of the SGRB FR, the main focus of our paper, with and without the LE, which is a $<$ 3 $\sigma$ effect.

\section{The sample selection}
\label{sample}

\begin{figure}
\includegraphics[width=250pt]{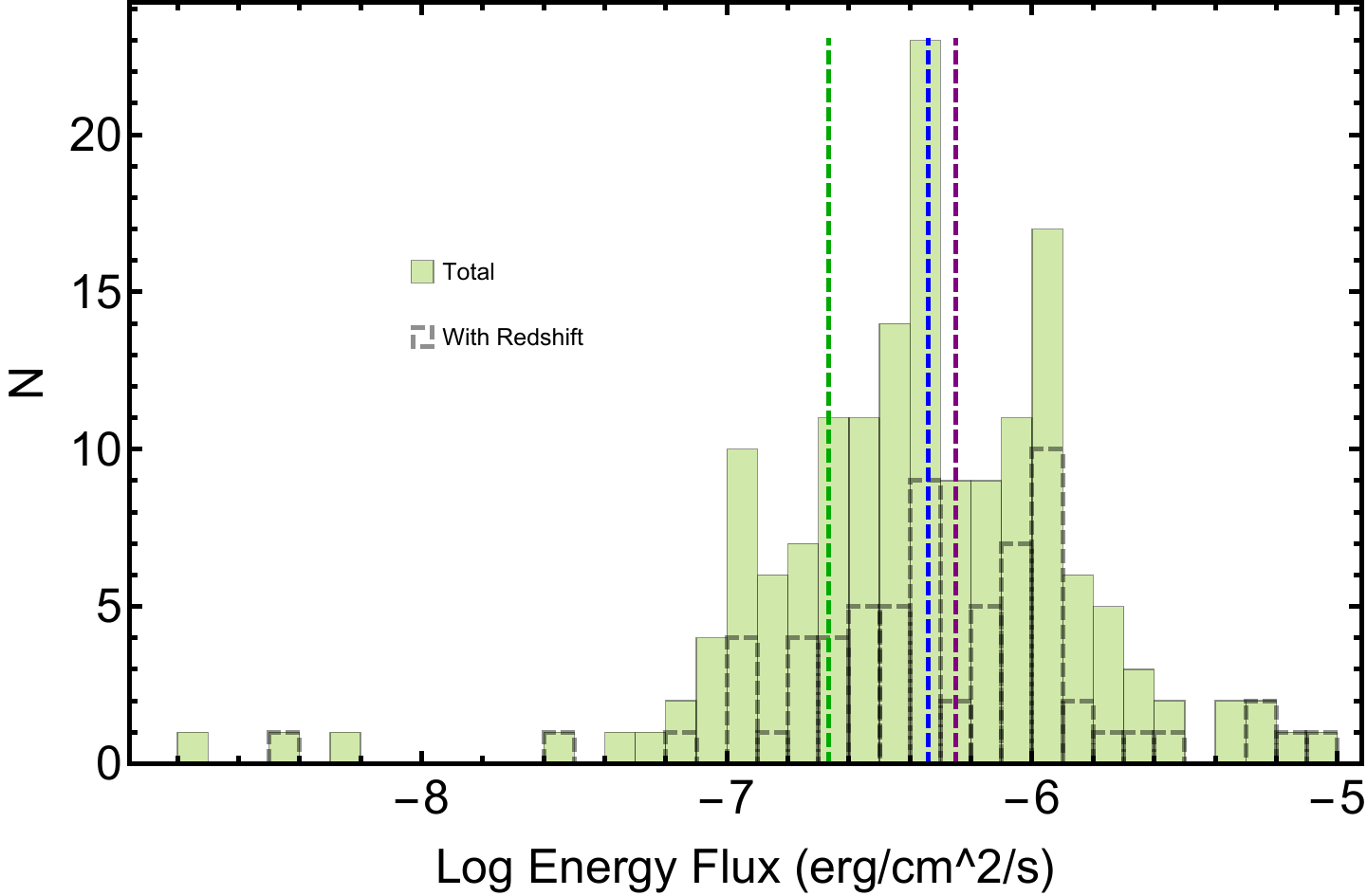}
\includegraphics[width=250pt]{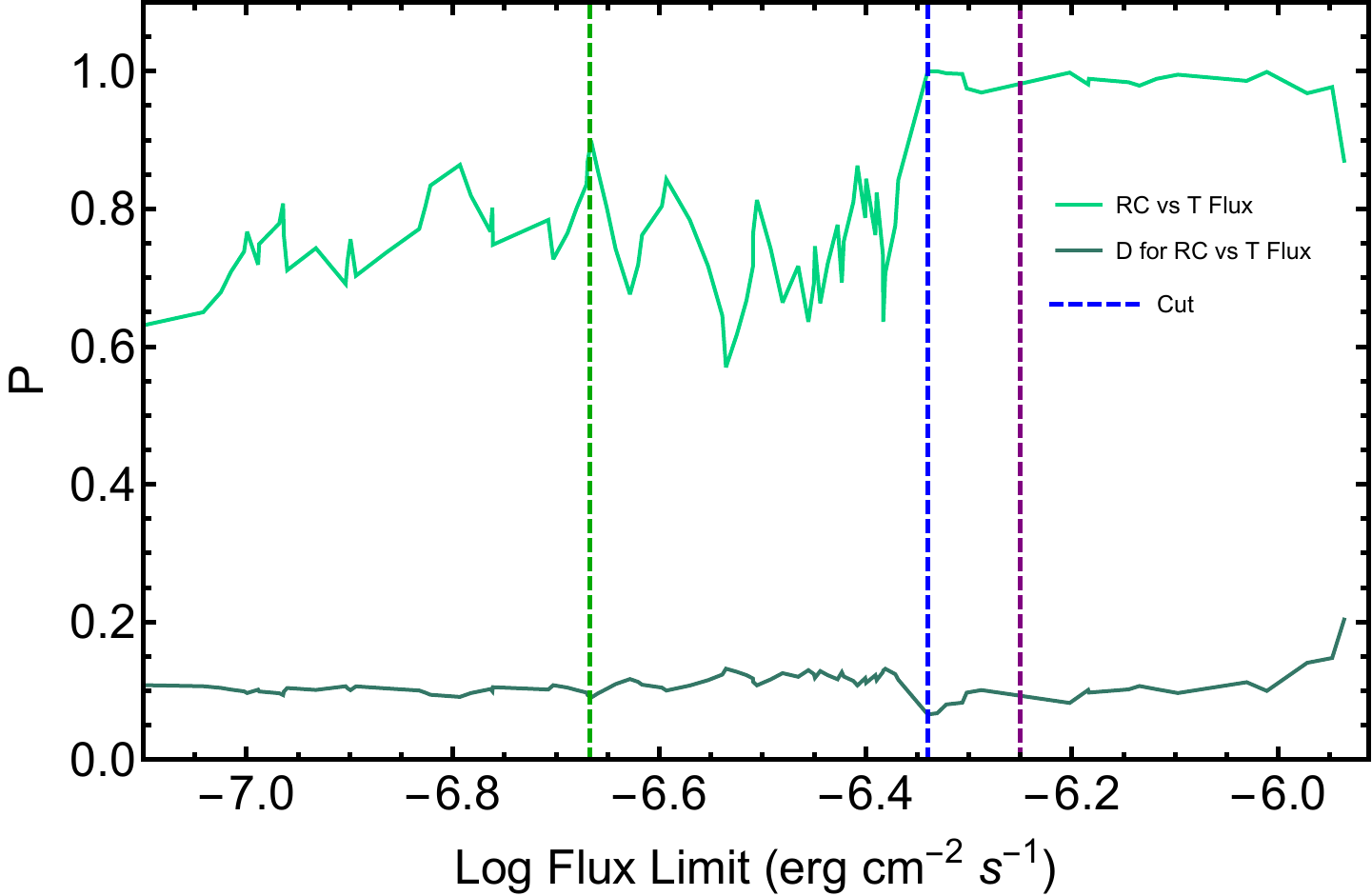}
\caption{Left panel: Histogram of the differential distribution of the fluxes of the 162  SGRBs and SEE GRBs (green filled bars), {\it the ``parent sample",} and the 68 with redshift (dashed black bars).  Right panel: The probability ($p$-value) as a function of the flux limit that  sub-samples with redshift are drawn from the parent sample obtained by KS test (upper bright green curve). The lower darker green line shows the maximum distance between the two cumulative distribution used in the KS test  to obtain the $p$ values (see examples in Figure 2). The vertical dashed green, blue  and purple  lines, in both panels show the flux limits of $\log f_{\rm lim}=-6.67$, $\log f_{\rm lim}=-6.34$ and $\log f_{\rm lim}=-6.25$ (in units of erg cm$^{-2}$ s$^{-1}$) with $p$-values of $0.9$, $1$ and $1$, respectively.} 
\label{Fhist}
\end{figure}

The Neil Gehrels Swift Observatory ({\it Swift}; \cite{G}) allows the rapid follow up (X-ray, optical/UV) observation after detection of the prompt emission.
As of June 2019, {\it Swift} BAT instrument has observed 1309 GRBs (1190 LGRBs and 162 SGRBs and SEEs) with a  given prompt flux limit. Of these, 472  have measured redshifts (339 LGRBs, and 68 SGRBs and SEEs). The observational selection criteria of the samples with redshifts is not well defined because redshift measurements are complicated involving localization by XRT and optical/UV follow up. 
%on board and/or ground based observations.
Therefore, there is a great difficulty in determining a well-defined flux limit, especially for samples with redshift. Swift has many triggering criteria for GRB detection in general  (Howell et al. 2014). To account for this problem, Lien at al. (2014) carry out extensive simulations to determine a  detection efficiency function that can be used in the determination of the LF. However, these simulations are based on characteristics that are more appropriate for Long GRBs.  Carrying out similar simulations for SGRBs would be useful, but is beyond the scope of this paper.
To overcome this difficulty we have extensively searched in the \textit{Swift} data bases, and have identified a complete sample of 162 SGRBs+SEEs with known peak fluxes, referred as the ``parent sample". This  sample is defined ``complete" or ``reliable" in the sense that we have all the information about the peak flux and the spectral features; it is the most comprehensive sample in the literature from December 2005 until June 2019.
From \cite{NB} we find a subsample of 68 SGRBs+SEEs with known redshifts (27 of which are SEE with redshift and are listed in Table 1). Figure \ref{Fhist} compares the differential flux distributions of the parent sample and the subsample with redshift.

\begin{figure}
\includegraphics[width=250pt,height=180pt]{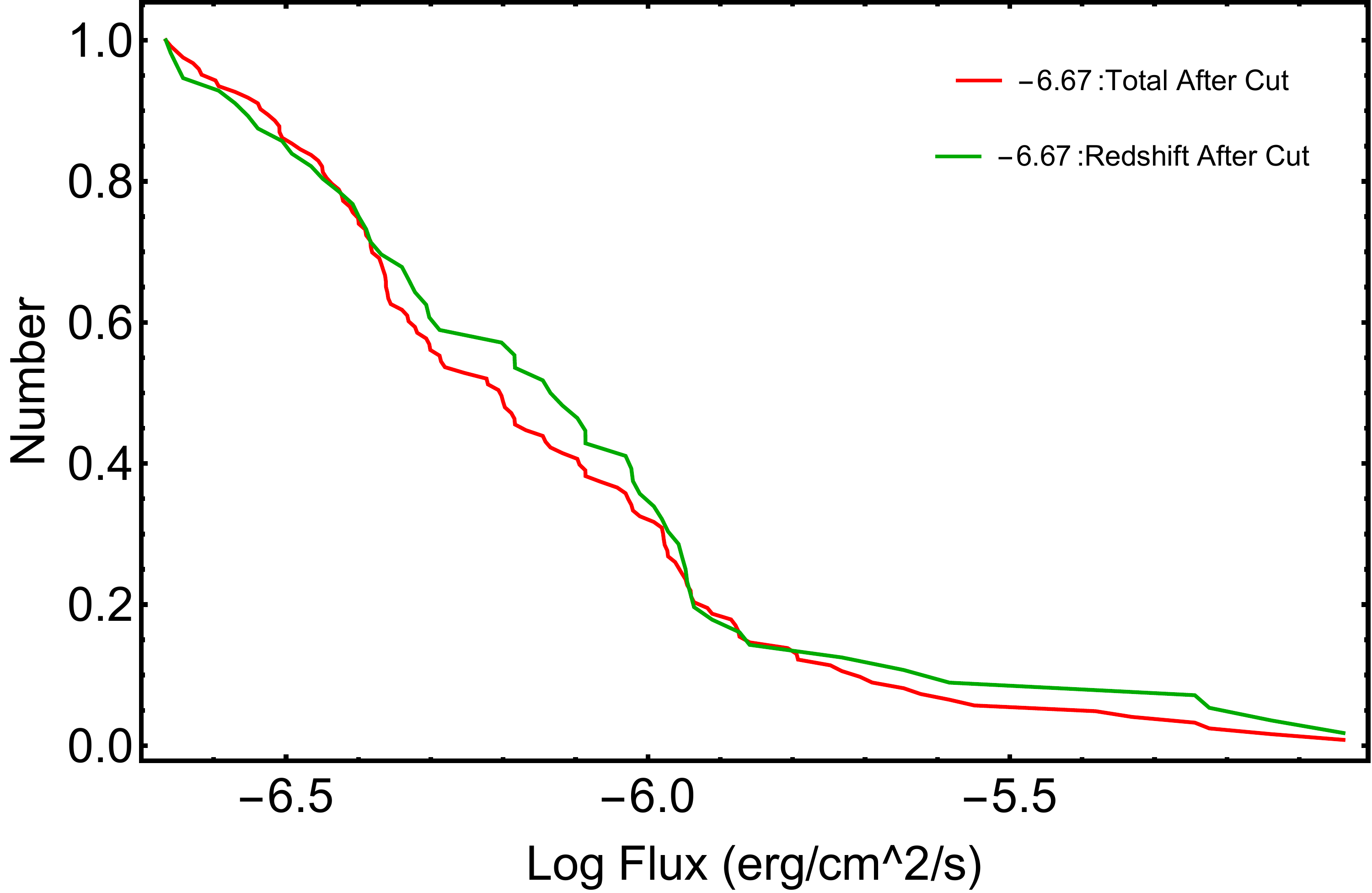}
\includegraphics[width=250pt,height=180pt]{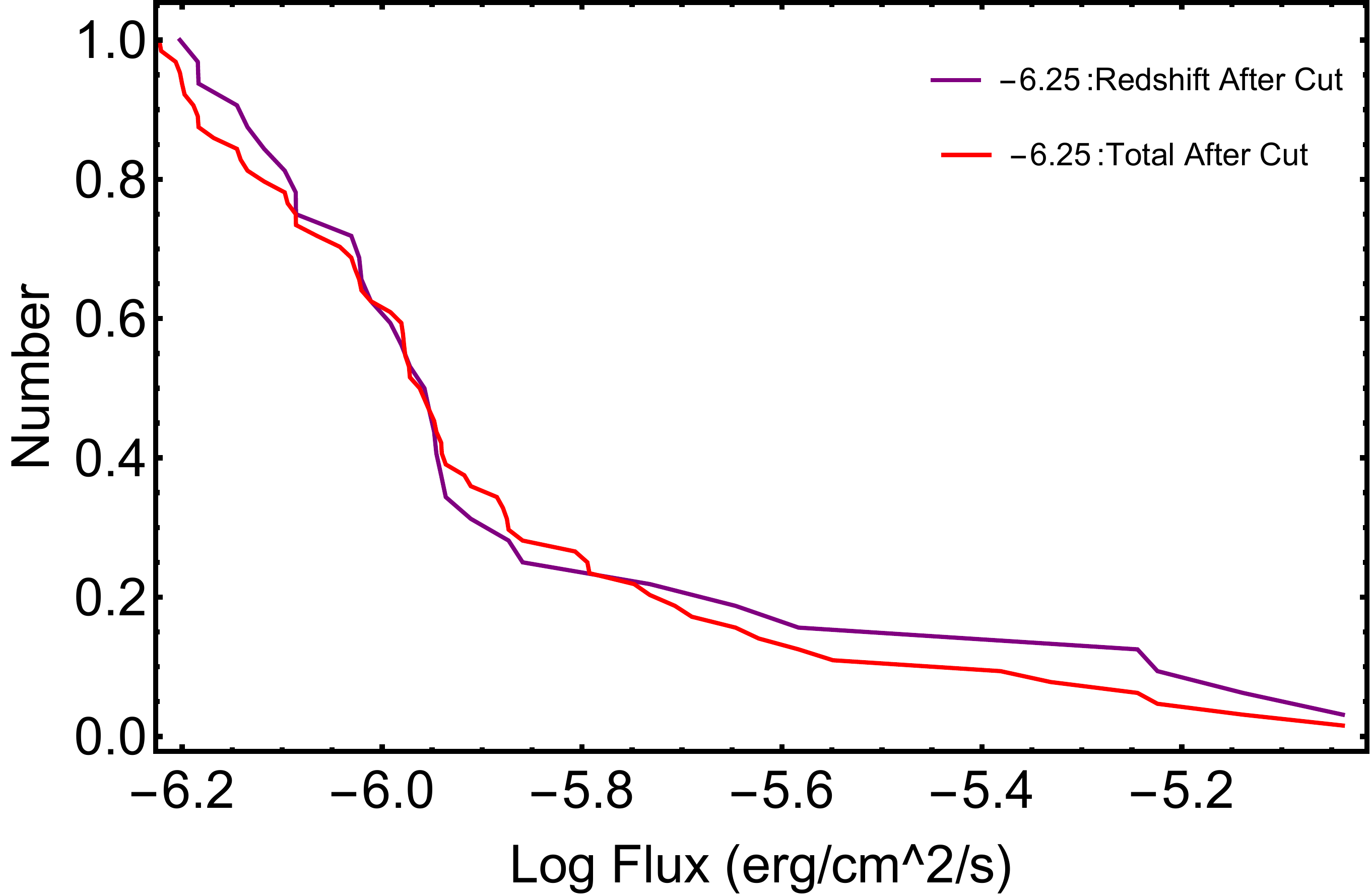}
\includegraphics[width=250pt,height=200pt]{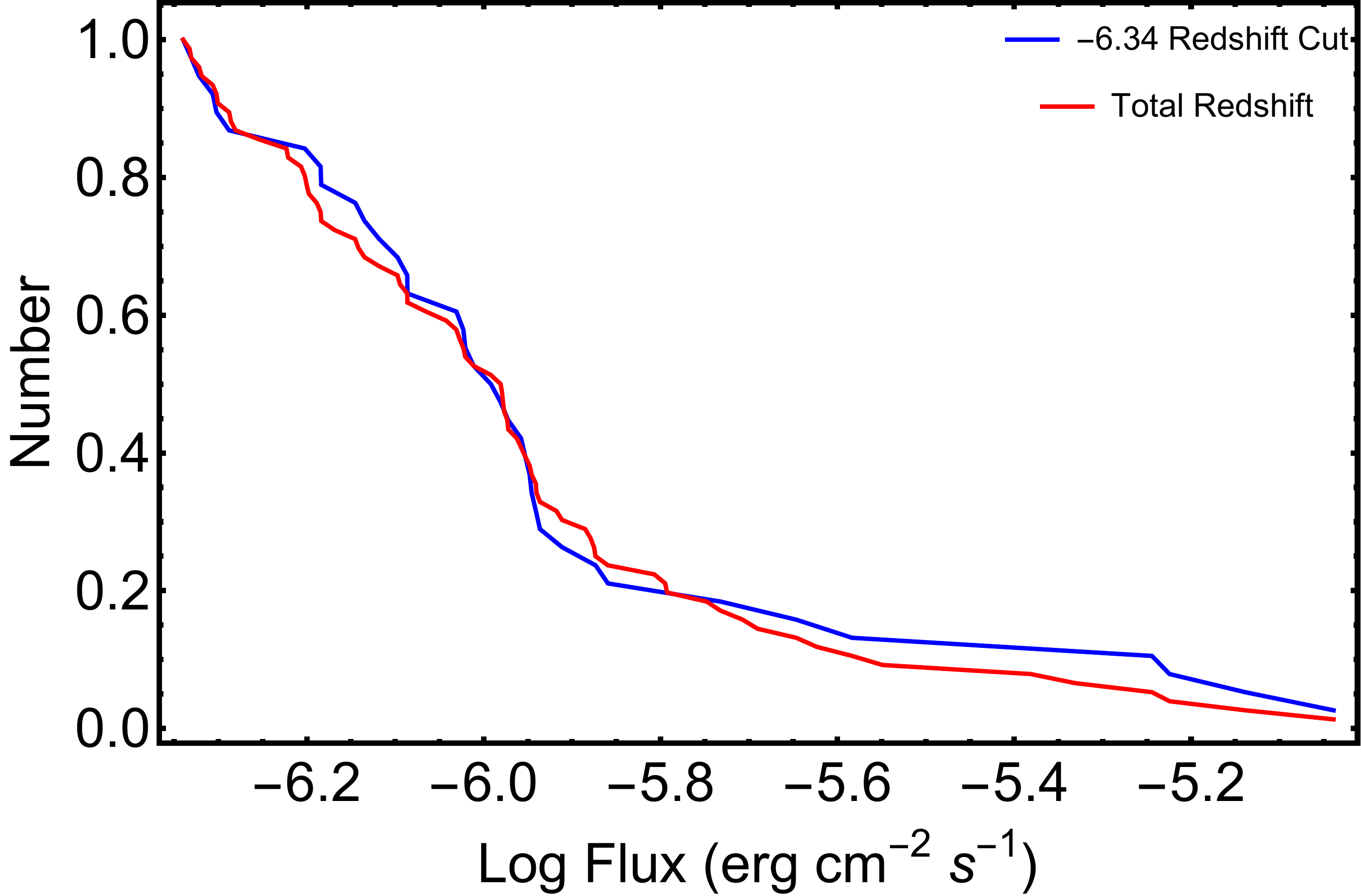}
\includegraphics[width=260pt,height=205pt]{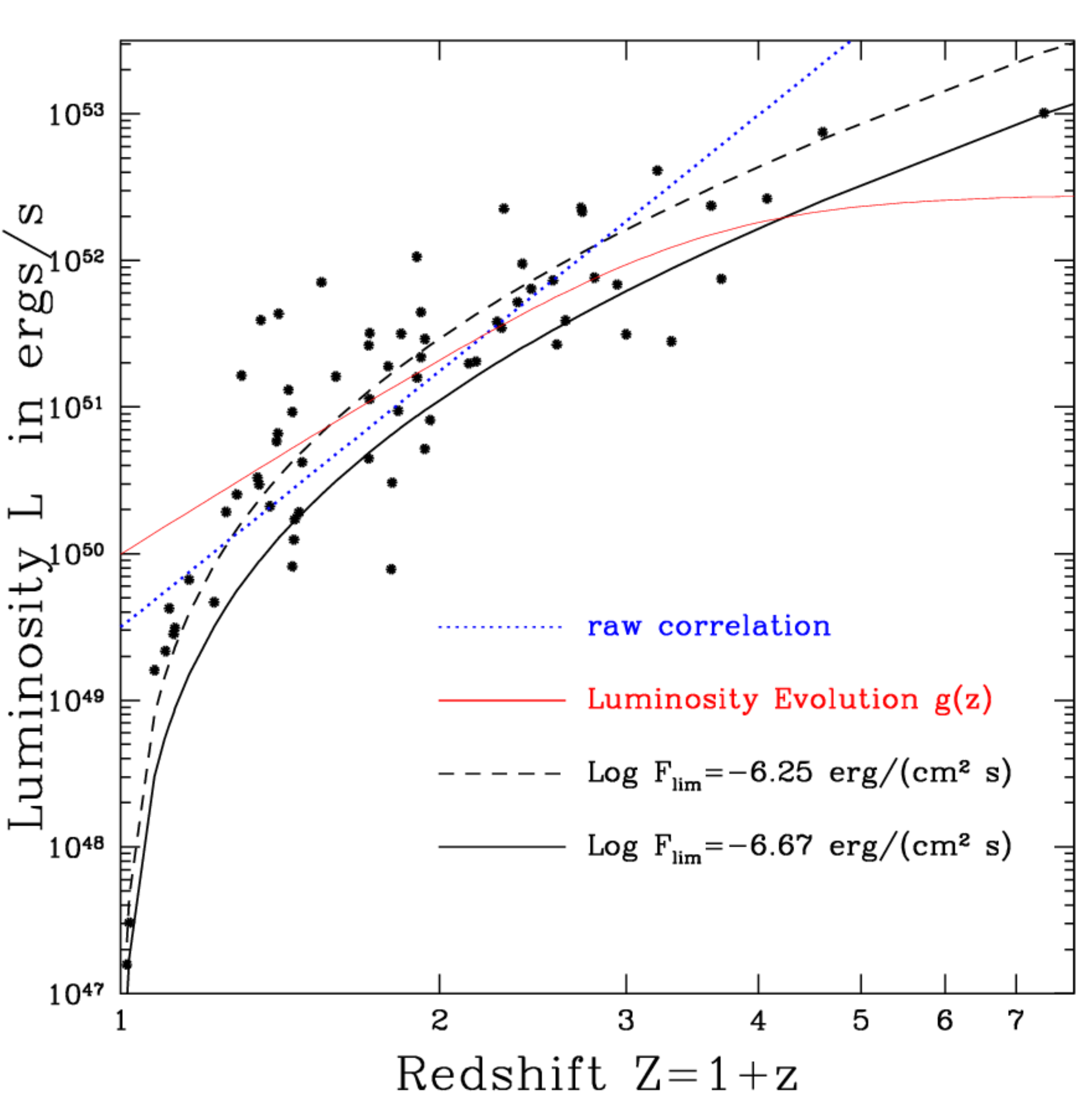}
\caption{Normalized cumulative distribution of the parent (red lines) and sub-samples with redshift (green, purple and blue lines)  used in the  K-S test for the three samples with $\log f_{\rm lim}=-6.67$ (upper-left), $-6.34$ (lower-left), and $-6.25$ (upper-right). Right lower panel shows the luminosity versus redshift distribution of all SGRBs and SEEs with known redshifts. The blue dotted line shows the raw correlation between luminosity and redshift, part of which is due to the truncation of the data caused by observational selection effects. The dashed  and solid black lines show the truncation boundaries, $L_{\rm min}(z)$, obtained for the cuts at $\log F_{lim}=-6.34$ and $-6.67$ erg $cm^{-2} s^{-1}$, respectively. The red curve shows the intrinsic correlation  obtained using the procedures described in \S 3.} 
\label{cumfluxdist}
\end{figure}

As expected, the fraction of sources with redshifts decreases with decreasing flux, from 0.53 for $f>f_c$ to 0.31  for $f_c>f>f_{\rm min}$, with $\log f_c=-6.3$ and $\log f_{\rm min}=-7.3$ (all fluxes hereafter will be in units of erg cm$^{-2}$ s$^{-1}$).
We then use the Kolmogorov-Smirnov (KS) test to determine the probability, $p$, that sub-samples with redshift are drawn from the parent sample, as a function of increasing flux limit starting with $\log f_{\rm min}=-7.3$. As shown in the right panel of Figure \ref{Fhist} 
 the $p-$value fluctuates between $0.6$ to $0.9$, eventually reaching a plateau with $p\simeq 1$  for $\log f>-6.34$. To show  the dependence of our results on the flux limit, we analyzed three samples: one with flux limit $\log f_{\rm lim}=-6.25$ well above the fluctuating part related to the probability that the samples are drawn by the same parent distribution (see left panel of Fig. 1, purple line) and one with $\log f_{\rm lim}=-6.34$ at the start of the plateau (both with $p=1$), and a third larger sample with $\log f_{\rm lim}=-6.67$, where there is a peak (with $p=0.9$). (The three limits are shown by the vertical dashed purple, blue, and green lines in Fig.~\ref{Fhist}.) The first two samples show very similar results (see Figures 2 and 3). Thus, in what follows we present results on the LF and FR for the larger sample with $\log f_{\rm lim}=-6.34$ (our ``Sample 1") consisting of 32 SGRBs with redshift, and for ``Sample 2" with $\log f_{\rm lim}= -6.67$, consisting of 56 sources  with known redshifts (34 SGRBs and 22 SEEs). 
The normalized cumulative distributions of fluxes of the three parent samples and their respective sub-samples with $z$, used in the KS test, are shown on
%the upper left (the cut at $\log f_{lim}-6.67$), right panel (the cut at $\log f_{lim}-6.25$) and lower left (the cut at $\log f_{lim}-6.34$) 
Figure \ref{cumfluxdist}. The right bottom panel of Figure \ref{cumfluxdist} shows the luminosity vs redshift of all sources and the two curves (the dashed and solid lines show the luminosity truncation, $L_{\rm min}(z)$, obtained from Equation (\ref{Lz}), for $\log f_{lim}=-6.34$ and $\log f_{lim}=-6.67$, respectively).

%A histogram detailing the total sample versus the number of GRBs with redshift is shown in the upper panel of Figure 2 and the total number versus the total number after applying the cut in flux is shown in the middle panel of Figure 2. The KS test calculates probability based on maximum distance between cumulative distributions. We here show in the lower panel Figure 2 the cumulative distributions for the total GRB sample after applying the cut versus the total 68 GRBs with redshift.

%
%                                                
%----------------------------------------------------------------- 
%\begin{figure}
%\includegraphics[width=500pt]{probability_distribution_MARIA_1_October.pdf}
%\caption{The probability (that the subsample with redshift is drawn from the parent %sample) derived from the Kolmogorov-Smirnov test as a function of the the flux limit.
%The purple dashed line indicates the cut we chose at log Flux=-6.67 with $P\sim 0.9$.}
%\label{KStest}
%\end{figure}

%with the limiting luminosity corresponding to $\log f_{\tm lim}=-6.67$.  The log luminosity versus redshift distribution with the limiting luminosity corresponding to $\log f_{\tm lim}=-6.67$.} Right panel: The red box (top-left) shows the associated set  (containing sources with $Z_j<Z_i$ and $L_j>L_{\rm min,i}$) for the GRB identified by the red circle at a redshift $Z_i$ and luminosity limit $L_{\rm min,i}$. The cyan dots indicate points for which the luminosity has been computed by using a cutoff power law fitting for the prompt emission spectrum, while the purple dots adopts the simple power law.}

%-----------------------------------------------------------------
\section{Results}
\label{results}

We determine the shape and evolution of the luminosity function $\Psi(L,Z)$ in Equation (\ref{LF}), using the observed $L-Z$ diagram corrected for biases introduced by the truncation in the samples, shown in the right panel of Figure \ref{cumfluxdist}.

%with luminosities (and their thresholds) calculated using  Equation (\ref{Lz}).
%Right panel of Figure 2 shows the $L-Z$ scatter diagram and the truncation boundary for $\log %f_{\tm lim}=-6.67$.

%\begin{figure}
%\includegraphics[width=500pt]{Lumi_lim_MARIA_1_October.pdf}
%\caption{The log luminosity versus redshift distribution with the limiting luminosity corresponding to $\log f_{\tm lim}=-6.67$. The red box (top-left) shows the associated set  (containin sources with $Z_j<Z_i$ and $L_j>L_{\rm min,i}$) for the GRB identified by the red circle at a redshift $Z_i$ and luminosity limit $L_{\rm min,i}$.}
%\label{LZdist}
%\end{figure}
\subsection{Luminosity Evolution}
\label{sec:LE}

\begin{figure}
\centering
\includegraphics[width=500pt]{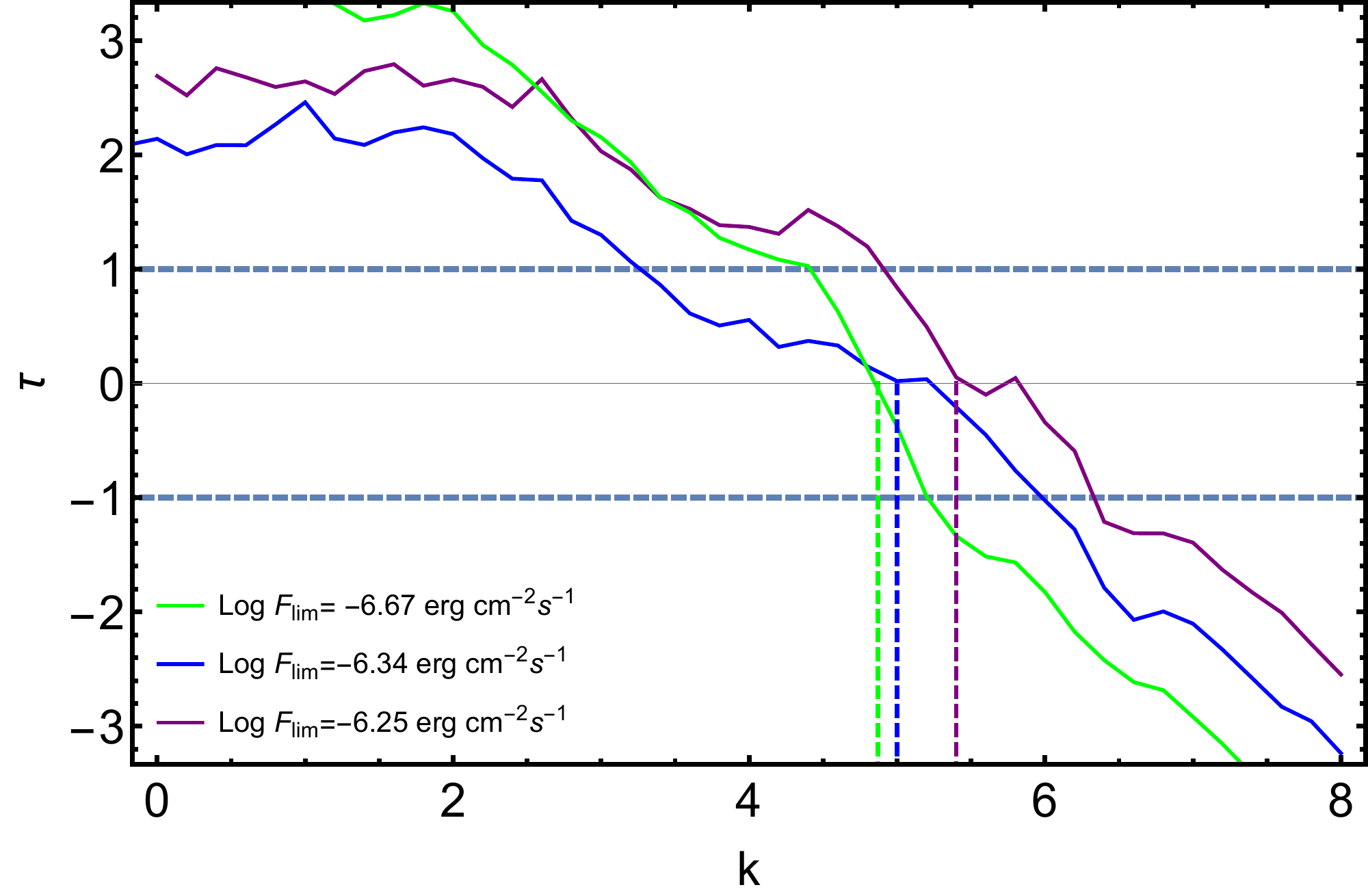}
\caption{Test statistic $\tau$ versus luminosity evolution index $k$, defined in equation \ref{gofz} with vertical lines giving the best values of $k$  that yields a local luminosity $L_0=L/g_k(Z)$ independent of (or uncorrelated with) redshift for the three samples with log flux limits of $-6.25$ (brown), $-6.34$ (blue) and $-6.67$ (green). The vertical dotted lines showing the best  (and $1 \sigma)$) values of $k=5.4^{+0.93}_{-0.5}, 5.0^{+1.0}_{-1.7}$ and $4.8^{+0.5}_{-0.5}$, respectively.}
\label{tau}
\end{figure}

As indicated in the right panel of Figure 2 (blue line),  the luminosity and redshift are highly correlated, but part of this correlation is due to truncation shown by the limiting curves. We use the \cite{EP} method to correct for this bias with a modified Kendell's $\tau$ statistics. We find a value $\tau$ $< 3$ indicating  a $\sim$ 3 $\sigma$ evidence for an intrinsic correlation or LE for {\bf all three} samples.  
We adopt {\bf a commonly used} single parameter evolutionary function {\bf $Z^k$ with slight modification}:
\begin{equation}
g_k(Z)=\frac{Z^k}{1+(Z/Z_c)^k}.
\label{gofz}
\end{equation}
 The denominator has  been added to reduce the rate of evolution at high redshifts (here chosen as $Z_c=3.5$) where the cosmic expansion time scale is reduced considerably from its current ($Z\leq 2$) value. This function has been proven very useful for  studies of high redshift  AGNs and GRBs (Singal et al. 2011, Petrosian et al. 2015, Dainotti et al. 2013, 2015b, 2017a).
 As demonstrated in Dainotti et al. (2015b) the difference between the results based on the simple function $g_k(Z)=Z^k$ and those based on Equation (\ref{gofz}) is $<2 \sigma$.
The variation of $\tau$ with $k$  is shown in Figure \ref{tau}, giving the best value of $k$ (when $\tau=0$) and its 1 $\sigma$ range of uncertainty (given by $|\tau| \leq 1$) of  $k=5.4^{+0.93}_{-0.5}, 5.0^{+1.0}_{-1.7}$ and $4.8^{+0.5}_{-0.5}$ for $\log f_{\rm lim}=-6.25, -6.34$ and $-6.67$, respectively. Thus, regardless of the flux limit the three values of $k$ appear to be strong (but $< 3$ $\sigma$) evidence of LE with
 $g_k(Z)\propto Z^{\sim 5}$ for $Z \le 3$. 
 
 Similar, but slightly slower LE was found for long GRBs (see. e.g.~Petrosian et al. 2015). However, if one underestimates $f_{\rm lim}$, one would obtain stronger evolution eventually reaching the maximum obtained from the raw data ignoring the effects of the truncation (i.e.~$f_{\rm lim}\rightarrow 0$). Recently,  Bryant et al. (arXiv:2010.02935v1) demonstrated this effect with simulations based on LGRBs characteristics. However, the three samples with different flux limits show very similar results, thus proving that this objection does not apply here. This effect comes into play when the truncation curve falls below most of the points, which, as evident in the right lower panel of Figure 2, is not the case here. On the other hand, if SGRBs progenitors are merging compact stars, it is not clear why such well defined events would depend strongly on the cosmological epoch of their occurrence. However, since we have meager observations on the generation of the electromagnetic radiation of the so-called kilonovae produced during such mergers, the existence of a LE cannot be ruled out. Nevertheless, because (i) the evidence is less than 3$\sigma$, (ii) there are uncertainties about the flux threshold, and (iii) there may be theoretical arguments against it,  we evaluate the FRE of the SGRBs, our main focus here, with and without inclusion of the LE.

\subsection{Luminosity function and Rate Evolution}
\label{Lfandrho}
%Note that if $\bar \alpha_e \noteq 0$ the K-correction modifies the evolution index to $k+\bar \alpha_e$.
%which represents the degree of correlation for the entire sample with proper accounting for the data truncation.
%The Kendall $\tau$ we use from the \cite{EP} method tests the independence of variables in a truncated data. Instead of calculating the ranks $R_i$ of each data point among all observed objects, which is customary for data without truncation, this method rank include only the point's `associated set. The `associated sets' include all objects that could have been observed given the observational limits. 

%Using $\tau$ correlation we find the luminosity evolution, see Figure 3.
%In this figure we plot with dashed vertical lines the . This uncertainty range regards the best fit value that represent the removal of evolution at $k=4.87^{+0.45}_{-1.24}$.

%For $\tau=0$ we obtain $k=4.87$ which establishes independence between the luminosity, L, and redshift, because the intrinsic luminosity, $L^{'}= L/g(z)$. We compute the cumulative luminosity function $\phi(L)$  and the cumulative rate evolution $\dot \sigma(z)$ shown in Figure 4 upper and lowe panel, respectively.

%Upper panel: Histograms of the cumulative local luminosity function $\Psi(L_0)$ (black crosses) obtained by EPL method. The red circles give the raw cumulative counts $N>L_0)$. The lines represent a fourth order polynomial fits used to obtain the differential distribution $\psi(L_0)$. Lower panel: Similar results shows the logarithm of the cumulative rate evolution $\dot \sigma(z)$ vs $\log Z$, and the raw counts $N(>Z)$ (red circles and the corrected rate (green triangles). The purple crosses give the corrected rate ignoring the luminosity evolution i.e.~assuming $k=0$ or  $g(Z)=1$. The curves show fifth order polynomial fits. Note that the results for $Z>3$ have high uncertainties because of the small numbers

\begin{figure}
\noindent
  \makebox[\textwidth]{\includegraphics[width=600pt]{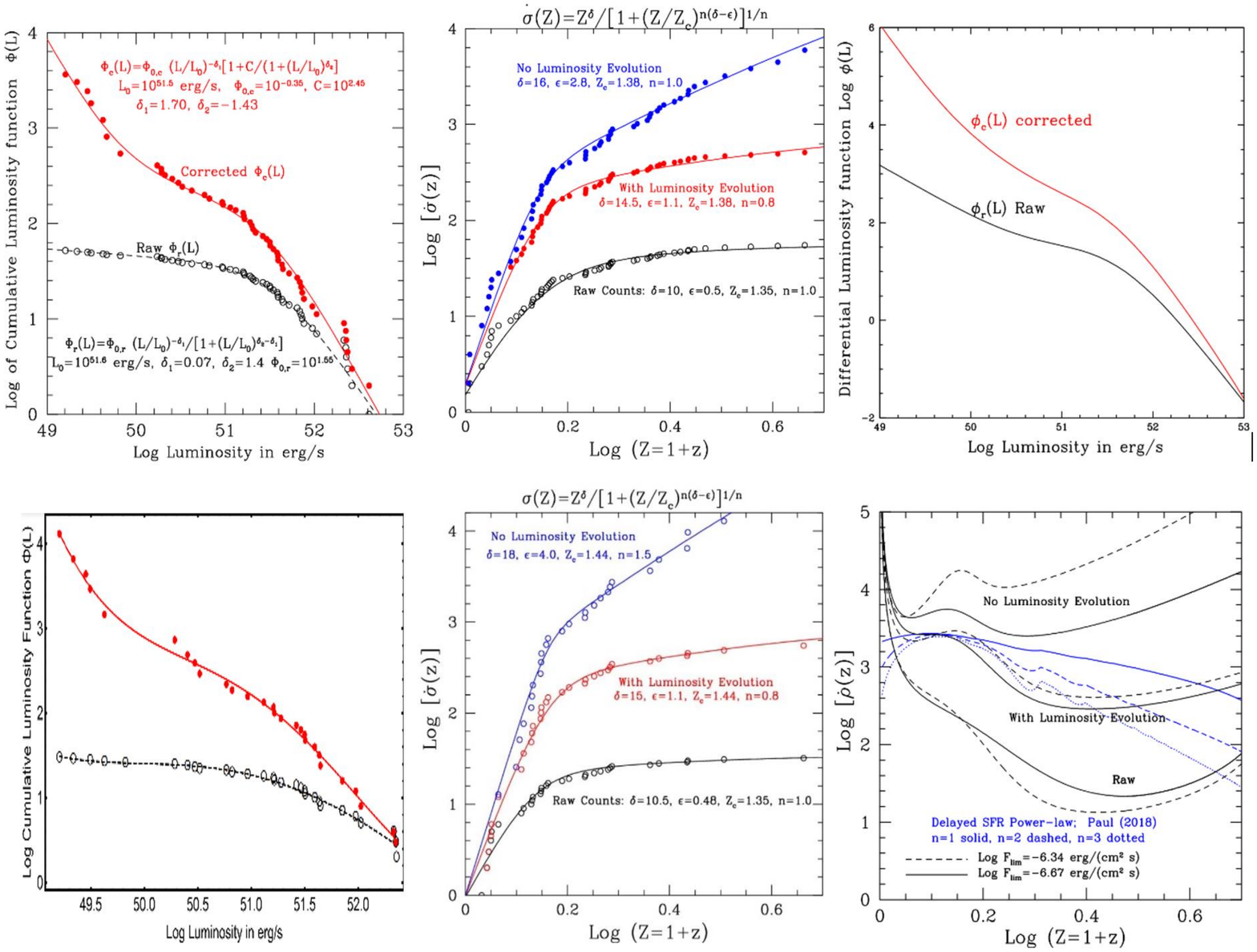}}
\caption{The cumulative LF, $\Phi(L_0),$ is shown in left panel. In the left panel raw data are shown with black points, while the points in red use the EPL method with no luminosity evolution. The density RE, ${\dot \sigma}(Z)$, is shown in middle panel for Sample 2 (upper panel, $\log F_{\rm lim}=-6.67$), and Sample 1 (lower panel, $\log F_{\rm lim}=-6.34$). In the middle panel the  black points show the raw counts $N(>L_0)$ and $N(<Z)$ (uncorrected for truncation). The red points are obtained by the EP-L non-parametric method correcting for truncation and LE. The blue points in the middle panels ignore LE (i.e.~set $k=0$). The lines on the left represent (identical) power-law fits with two breaks, with their forms and parameters given inside the top panel. The black, red, and blue lines in the middle panels are obtained from a simple broken law form given above the panels with parameters given inside the panels. There are slight differences between the values of the parameters for the two samples. The analytic functions  are used to obtain the differential distributions $\psi(L_0)=-d \Phi (L_0)/dL_0$ shown on the right upper panel (same for both samples), and ${\dot \rho}(Z)=Z(d{\dot \sigma}(Z)/dZ)/(dV/dZ)$ shown on the lower panel for Sample 1 (dashed) and 2 (solid) (raw, with LE, and without LE). The ${\dot\sigma}$ curve for the no LE case is steeper in the lower middle panel giving rise to higher density rate evolution, ${\dot \rho}(z)$, shown by the dashed lines in lower right panel.
On this panel, we also show three curves (long-dashed, dashed and dotted) taken from the Paul (2018) calculation of the delayed SFR with a power-law distribution of the delay times with indexes 1, 2 and 3, respectively.}%\includegraphics[width=250pt]{PhiSEECor.pdf}

\label{VAhe}
\end{figure}

Having established the independence of luminosity $L_0$ and $Z$ we can then proceed to obtain their distribution following the steps of EP-L method. This method gives non-parametric histograms of the cumulative distributions 

\begin{equation}
\Phi(L_0)=\int_{L_0}^\infty\phi(L'_0)dL'_0\,\,\,\, {\rm and}\,\,\,\,  \dot \sigma(Z)=\int_1^Z{\dot \rho}(Z')\left({dV\over dZ'}\right){dZ'\over Z'},
\end{equation}
derivatives of which give the differential distributions. Here $V(Z)$ is the
co-moving volume up to $Z$. Upper and lower left and middle panels of Figure \ref{VAhe} 
show the two cumulative distributions we obtain with FP-L method, compared to raw  cumulative counts (not corrected for the truncation), for the two samples ($\log f_{\rm lim}= -6.67$, upper and -6.34 lower), respectively. In the case of rate evolution, shown in the middle panels, we show the corrected values including LE ($k=5.0$, red) and without LE ($k=0$, blue). The two samples give very similar results. Thus, very similar results will be given also for the sample at $\log f_{\rm lim}=-6.25$, not shown to avoid cluttering the pictures. Non-parametric derivatives of these histograms can be obtained directly from the data as well. However, given that the cumulative distributions are somewhat noisy, we fit the cumulative distributions with analytic forms (broken power-laws). The forms and the values of the parameters are shown inside the panels of Figure \ref{VAhe}. For the LF the two samples are fit by exactly the same functions, but for cumulative rate evolution the results are slightly different as a comparison with fitted parameter values on the upper and lower panels would indicate. 

The differential LF, $\psi(L_0)=-d\Phi(L_0)/dL_0$, obtained from the fitted forms (identical for both samples) is shown on the upper right panel of Figure \ref{VAhe}, and the differential density rate evolution,
${\dot\rho}(z)=Z[d{\dot\sigma}(z)/dz]/[dV(z)/dz]$,
for both samples, with and without LE, are shown on the lower  right panel  of this figure showing some range of possibilities and thus uncertainties between the two samples, with and without accounting for LE. We have also plotted  curves for delayed SFR with a power law delay distribution with three indexes taken from Paul (2018) normalized at their peak values. We can see from the right bottom panel of Fig. 4 that there is agreement with Paul (2018) for ranges of redshift between $z=0.10$ to $z \sim 1$.
%\textcolour{purple}{In the right bottom panel of figure Figure \ref{VAhe} we have shown also the Wanderman & Piran log density rate evolution for comparison with our results. We here stress that the results of Wandermann & Piran (2015) are different and we acknowledge this difference due to the fact that the density rate evolution derived by us is independent from the assumption derived from forward fitting method and from the assumption of a given formation rate, but they are derived directly from the data.}

%\begin{figure}
%\includegraphics[width=\linewidth]{rhovsz.pdf}
%\caption{Upper panel: The differential local luminosity function fitted to a broken power law. Bottom panel: Three co-moving density rate evolution obtained from the corresponding three fits to the cumulative distributions shown in bottom panel of  Figure %\ref{cumdists}.}
%\label{diffdists}
%\end{figure}

\section{Summary and Discussion}
\label{summary}
From a total sample of 162 GRBs (118 SGRBs and  44 SEE GRBs) with known redshifts and spectra we have selected three sub-samples with flux limits of $\log f_{\rm lim}=-6.25, 6.34$ and -6.67 consisting of 32, 34 SGRBs and 56 GRBs (34 SGRBs and 22 SEE GRBs). These samples according to the KS tests have probabilities of 1, 1 and 0.9, respectively, that are drawn from the ``parent samples" consisting of all sources (with or without redshift) with the same flux limits.
Using the non-parametric combined EP-L methods we obtain the following results, which are very similar and compatible with 1 $\sigma$ for the three samples.

 \begin{enumerate}
     
 \item 

For the three samples we find very similar evidence of a $\leq 3\sigma$ LE, $L(Z)\propto Z^{\sim 5}$ at low $z$ tending to constant value at $z > 2.5$, that is much stronger than  $Z^{\sim 3}$ obtained for LGRBs. 

\item

After correction for the LE, we derive the cumulative LF, and its derivative the differential LF, which unlike those of LGRBs, that can be fit with a simple broken power-law, shows  steepening at (local) luminosities $L_0<10^{50}$ erg s$^{-1}$ with an index equal to that at high luminosities of $L>10^{51.6}$ erg s$^{-1}$, possibly caused by an excess of low redshift sources. Most FF methods (see, e.g. Wanderman and Piran, 2015) assume a priori a simple broken power law form for the LF of SGRBs. Our results,  obtained directly from the  data non-parametrically and without any assumptions, indicates that, unlike LGRBs,  a power-law with one break may not provide an adequate description of the LF of SGRBs.

\item With the same procedure we find the cumulative  and the differential co-moving density rate evolution with redshift. Here, we obtain the rate evolution with and without inclusion of the $<3 \sigma$ LE. In general, these rates decrease rapidly at low $z$, but flatten out at higher $z$. Inclusion of the LE yields a monotonically decreasing rate up to $z=2.5$. Beyond this, the rate increases slowly but this behavior is highly uncertain because of the small numbers of sources with $z>3$. At very low redshifts this rate is in disagreement  with the SFR (which increases rapidly as $Z^{2.7}$ up to $Z\sim 3$). 
This low redshift behavior is similar  to that found for LGRBs.  But unlike LGRBs, the SGRB rate disagrees with the standard SFR at high redshifts too. It should be noted that there are very few SGRBs at $z<0.1$  and $z>3$ making these portions of the rate more uncertain. On the other hand, the somewhat monotonic decrease, especially with LE, is what is  expected for events, such as merging compact  binaries, with considerable time delay relative to SFR (see, e.g. Wanderman \& Piran 2015 and Paul 2018). As shown in Fig.~4 (lower right), such expected rates agree with our results in mid-redshift range. 

\end{enumerate}

These results on cosmological distributions and the evolution of SGRBs are based on a well defined and relatively sizable ``reliable sample" with measured redshifts using powerful non-parametric and non-binning methods. 
In the future, we will repeat the same treatment to further constrain the density rate evolution and the luminosity function by increasing the sample size. This will be possible via the inclusion of GRBs observed by other instruments such as {\it Konus}-Wind and {\it Fermi}-GBM. The non-parametric methods used here are ideally suited for combining data from different instruments with different energy bands and selection criteria. With more data and more accurate determination of the FR of SGRBs it may be possible to constrain the parameters of the delayed SFR models.

\begin{acknowledgements}
L. Bowden acknowledges the support by the U.S. Department of Energy, Office of Science, Office of Workforce Development for Teachers and Scientists (WDTS) under the Science Undergraduate Laboratory Internships Program (SULI). 

M.G.D. acknowledges the support of the American Astronomical Society Chretienne Fellowship that allow her to be hosted at Stanford in 2018-2019 and NAOJ Division of Science for the current support. M.G.D. is grateful to Dr. Cuellar for managing the SULI program. We are grateful to Hannah Yasin Hashai for useful discussion about our results.
\end{acknowledgements}


\begin{appendix} %First appendix
\section{Data}
% In the appendices do not forget to put the counter of the table 
% as an option
%\longtab[1]{
%\begin{longtable}
\begin{table}
 \centering
 \caption{SEE GRBs with known redshift. Epeak assumes cutoff power law unless otherwise noted.}
 \begin{tabular}{cccccccc} % four columns, alignment for each
  \hline
  Gamma Ray & Alpha & Epeak (keV) & Energy Flux & T90 & SEE-Source & z & z-source \\
  Burst &  &  & 15-150 keV (erg/$cm^2$/s) & (s) &  &  &  \\
  \hline
  050724A & -0.78 & 77.81430 & 9.53E-07 & 96 & REF[7,9,10] & 0.2570 & G/REF[16]\\
  050911 & -0.95 & pl & 2.69E-07 & 16.2 & REF[7,8,15] & 1.1650 & REF[3]\\
  051016B & 2.71 & 44.03600 & 1.17E-07 & 4 & REF[9,11,14] & 0.9364 & G/REF[16]\\
  051227 & -1.03 & pl & 1.96E-07 & 114.6 & REF[5,6,11] & 0.7140 & G\\
  060306 & -0.91 & 140.58500 & 4.67E-07 & 61.2 & REF[4,7] & 1.5590 & G\\
  060607A & 0.39 & 227.35400 & 3.22E-07 & 102.2 & REF[7] & 3.0749 & G/REF[16]\\
  060614 & -1.03 & 230.29500 & 7.62E-07 & 108.7 & REF[5,7,8] & 0.1250 & G/REF[16]\\
  060814 & -0.46 & 152.90200 & 9.31E-07 & 145.3 & REF[7] & 0.8400 & REF[16]\\
  060912A & -1.82 & 590.16700 & 6.55E-07 & 5 & REF[1] & 0.9370 & G/REF[16]\\
  061006 & 0.04 & 305.73800 & 1.85E-06 & 129.9 & REF[5,7,9] & 0.4400 & G\\
  061021 & -0.89 & 461.80200 & 8.20E-07 & 46.2 & REF[4,8] & 0.3463 & G\\
  061210 & 0.03 & 256.95100 & 7.26E-06 & 85.3 & REF[7,8,9] & 0.4095 & G/REF[16]\\
  070223 & -1.03 & pl & 2.23E-07 & 88.5 & REF[7] & 1.6295 & G\\
  070506 & -1.19 & pl & 6.84E-08 & 4.3 & REF[9,14] & 2.3100 & REF[16]\\
  070714B & -0.70 & 9794.42000 & 1.05E-06 & 64 & REF[7,8,11] & 0.9200 & G/REF[16]\\
  080603B & -1.08 & pl & 1.26E-07 & 60 & REF[7] & 2.6900 & G/REF[16]\\
  080913 & 0.07 & 67.87210 & 2.16E-07 & 8 & REF[2,3] & 6.4400 & G/REF[16]\\
  090530 & 0.71 & 64.77000 & 4.08E-07 & 48 & REF[7,14] & 1.2660 & G/REF[16]\\
  090927 & -0.94 & 851.06400 & 4.57E-07 & 2.2 & REF[9,14] & 1.3700 & REF[16]\\
  100704A & -0.77 & pl & 6.28E-07 & 197.5 & REF[7] & 3.6000 & REF[16]\\
  100814A & 0.34 & 171.65700 & 4.99E-07 & 174.5 & REF[7] & 1.4400 & G/REF[16]\\
  100816A & 0.70 & 108.27700 & 1.00E-07 & 2.9 & REF[7,14,15] & 0.8040 & G/REF[16]\\
  100906A & 0.90 & 78.45730 & 1.07E-06 & 114.4 & REF[7] & 1.7270 & G/REF[16]\\
  111005A & -1.42 & pl & 3.98E-07 & 26 & REF[13] & 0.0133 & G\\
  111228A & -1.38 & 214.95800 & 1.16E-06 & 101.2 & REF[9] & 0.7140 & G/REF[16]\\
  150424A & -0.11 & 998.74700 & 5.70E-06 & 91 & REF[11,12,15] & 0.3000 & REF[16]\\
  160410A & 0.77 & 197.81600 & 1.15E-06 & 8.2 & REF[11,14,15] & 1.7200 & G/REF[16]\\
  \hline
 \end{tabular}
\end{table}
%\end{longtable}
%}% End longtab
\end{appendix}
$\\$
REF[1] = Levan at al. 2007, MNRAS, 378, 143
$\\$
REF[2] = Ghirlanda et al. 2009, A\&A, 496, 585
$\\$
REF[3] = Zhang, Bing et al. 2009, ApJ, 703,1696–1724, 1.
$\\$
REF[4] = Minaev et al. 2010, AstL, 36, 707M
$\\$
REF[5] = Norris et al. 2010, ApJ, 717, 411
$\\$
REF[6] = Gompertz et al. 2013, MNRAS, 431, 1745
$\\$
REF[7] = Hu et al. 2014, ApJ, 789, 145
$\\$
REF[8] = Van Putten et al. 2014, MNRAS, 444, L58
$\\$
REF[9] = Kaneko et al. 2015, MNRAS, 452, 824
$\\$
REF[10] = Abbott et al. 2017, ApJL, 848, L13
$\\$
REF[11] = Gibson et al. 2017, MNRAS, 470, 4925
$\\$
REF[12] = Knust et al. 2017, A\&A, 607A, 84K
$\\$
REF[13] = Wang et al. 2017, ApJL, 851, L20.
$\\$
REF[14] = Anand et al. 2018, MNRAS, 481, 4332A
$\\$
REF[15] = Kagawa et al. 2019, ApJ, 877, 147K
$\\$
REF[16] = \url{https://swift.gsfc.nasa.gov/archive/grb_table/}
$\\$
G = \url{http://www.mpe.mpg.de/~jcg/grbgen.html}
\end{document}